\documentstyle[sprocl]{article}
\input{psfig}
\def\Journal#1#2#3#4{{#1} {\bf #2}, #3 (#4)}

\def\NPB{{\em Nucl. Phys.} B}
\def\PLB{{\em Phys. Lett.}  B}
\def\PRL{\em Phys. Rev. Lett.}
\def\PRD{{\em Phys. Rev.} D}

\def\st{\scriptstyle}

\def\be{\begin{equation}}
\def\ee{\end{equation}}
\def\bea{\begin{eqnarray}}
\def\eea{\end{eqnarray}}
\begin{document}

\title{$B^- \to \mu^- {\bar \nu}_\mu \gamma$ AND THE DETERMINATION
OF $f_B$ }

\author{ F. DE FAZIO}

\address{Dipartimento di Fisica dell'Universit\'a di Bari,\\
Istituto Nazionale di Fisica Nucleare, Sezione di Bari,\\
Via Amendola 173, 70126, Bari, ITALY}

\maketitle\abstracts{
We compute the rate of the decay $B^- \to \mu^- {\bar \nu}_\mu \gamma$ by 
a QCD relativistic potential model, 
with the result:
${\cal B}(B^- \to \mu^- {\bar \nu}_\mu \gamma)=0.9 \cdot 10^{-6}$. 
We also discuss how this decay mode can be used to access 
$f_B$.}
 
\section{Introduction}
The $B$ meson leptonic constant $f_B$, defined by 
$<0|{\bar q}\gamma^\mu \gamma_5 b|B(p)>=i \; f_B p^\mu $,
 plays a prime role in the physics of heavy-light quark systems. 
In principle, it can be obtained from the purely leptonic 
decays: $B^- \to \ell^- {\bar \nu}_\ell$, whose rates
are proportional to $f_B^2$.
However, the
helicity suppression, represented by
a factor $ (m_\ell/m_B)^2$, implies very low decay rates for 
$\ell=e,\mu$;
using $V_{ub}=3 \cdot 10^{-3}$ and $f_B=200\; MeV$ one predicts: 
${\cal B}(B^- \to e^- {\bar \nu}_e)\simeq 6.6 \; 10^{-12}$;
${\cal B}(B^- \to \mu^- {\bar \nu}_\mu)\simeq 2.8 \; 10^{-7}$ \footnote{
Experimental bounds \cite{cleo} are:
${\cal B}(B^- \to e^- {\bar \nu}_e)< 1.5 \; 10^{-5}$,
${\cal B}(B^- \to \mu^- {\bar \nu}_\mu)<2.1 \; 10^{-5}$.}. 
This problem is absent in the 
$\tau$ channel, which however presents identification difficulties.
One may ask whether $f_B$ could be obtained from other decays.
A candidate \cite{burdman,noi} is 
$B^- \to \mu^- {\bar \nu}_\mu \gamma$, since a third body
in the final state prevents helicity suppression.
Indeed,  using a relativistic potential model we predict
 that its branching ratio is higher 
than in the purely leptonic channel \cite{nois}; this gives us the possibility 
to access $f_B$ from this decay mode. 

\section{$B^- \to \mu^- {\bar \nu}_\mu \gamma $ in a Relativistic Potential 
Model}

\begin{figure}[ht]
\begin{center}
\mbox{\psfig{file=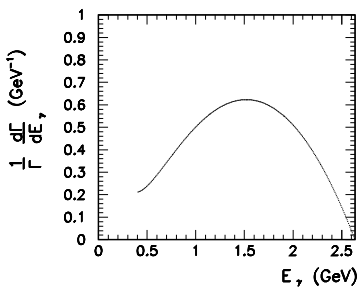,height=1.5in}}
\end{center}
\caption{Photon energy spectrum in the decay 
$B^- \to \mu^- {\bar \nu}_\mu \gamma$. \label{th251_fig1}}
\end{figure}

The amplitude of the decay 
$B^-(p) \to \mu^-(p_1) \; {\bar \nu}_\mu(p_2) \; \gamma(k, \epsilon)$ 
can be written as
${\cal A}(B^- \to \mu^- {\bar \nu}_\mu \gamma)
={G_F \over \sqrt{2}} V_{ub} \; \; \big( L^\mu \cdot \Pi_\mu \big)$,
\noindent where  
$L^\mu={\bar \mu}(p_1) \gamma^\mu (1-\gamma_5)\nu(p_2)$
is the weak leptonic current and
$\Pi_\mu = \Pi_{\mu \nu} \epsilon^{* \nu}$
 is the hadronic correlator:
\begin{equation}
\Pi_{\mu \nu}= i\; \int d^4 x 
e^{i q \cdot x} <0|T[ J_\mu(x) V_\nu(0) ]|B(p)>  \hskip 3 pt .  \label{pi} 
\end{equation}
\noindent 
In (\ref{pi})  $q=p_1+p_2$,
$J_\mu={\bar u} \gamma_\mu (1 -\gamma_5) b$
is the weak hadronic current and 
$V_\nu={2 \over 3} e\; {\bar u} \gamma_\nu u-
{1 \over 3} e \; {\bar b} \gamma_\nu b$ is the electromagnetic 
current containing  the coupling of the photon 
both to the light and the heavy quark.
The $B$ meson is described by a wave function 
$\psi_B({\vec k_1})$ 
 representing  the distribution of the heavy quark momentum ${\vec k}_1$
inside the hadron. 
In the $B$ rest frame it can be obtained 
by solving numerically a Salpeter equation which includes  
 relativistic effects in 
the kinematics \cite{pietroni}. 
Quark interaction is described by the Richardson 
 potential, reproducing  QCD phenomenology since it is linear for large 
distances, and therefore confining, and it behaves as $\alpha_s(r) \over r$ 
for small $r$, according to asymptotic freedom. Spin interaction can be 
neglected, being ${\cal O}(m_b^{-1})$. 
In this framework, we obtain \cite{nois}:
\begin{equation}
\Gamma(B^- \to \mu^- {\bar \nu}_\mu \gamma)= {G_F^2 |V_{ub}|^2 \over
3 (2 \pi)^3} \int_0^{m_b/2} dk^0 k^0 (m_b-2k^0)[|\Pi_{11}|^2 +
|\Pi_{12}|^2] \hskip 3 pt  \label{gamma} 
\end{equation}
where $\Pi_{11}$ and $\Pi_{12}$ derive from  (\ref{pi}) 
 \cite{nois}. Moreover, the bound 
$|{\vec k}_1| \le {m_B^2-m_u^2 \over 2 m_B}$ 
is obtained  
assuming that the $b$ quark has a running mass: 
$m_b^2({\vec k}_1)=m_B^2+m_u^2-2m_B ({\vec k}_1^2 + m_u^2)^{1/2}$, 
 and imposing $m_b^2 \ge 0$, an equation 
  stemming  from the 
requirement of energy conservation at quark level \cite{accmm}: 
$E_b + E_u=m_B$ 
($E_{b,u}=({\vec k}_1^2+m^2_{b,u})^{1/2}$).
Finally, we put a cut-off $\Lambda$ for small photon energies $k^0$  
 to avoid the unphysical divergences in (\ref{gamma}) for
$k^0 \to 0$, corresponding to no photon emitted in the final state
\footnote{The inclusion of radiative corrections would 
formally cancel the divergence; notice that only 
photon energies larger than $50 \; MeV$ 
are experimentally measurable.}.
Using $\Lambda=350 \; MeV$, we obtain \cite{nois}:
\begin{equation}
{\cal B}(B^- \to \mu^- {\bar \nu}_\mu \gamma) =0.9 \cdot 10^{-6}~~ \; 
\label{br}
\end{equation}
\noindent  a result that depends very slightly on $\Lambda$.
The computed photon spectrum (Fig. \ref{th251_fig1}) shows a 
peak around $1.5 \; GeV$.

\section{How to relate $f_B$ to $B^- \to \mu^- {\bar \nu}_\mu \gamma $ }
\begin{figure}[ht]
\begin{center}
\mbox{\psfig{file=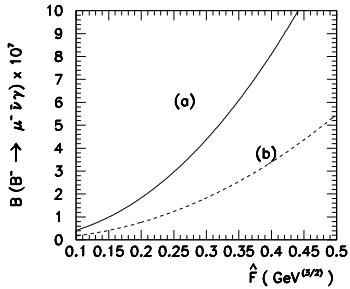,height=1.5in}}
\end{center}
\caption{The branching ratio
${\cal B}(B^- \to \mu^- {\bar \nu}_\mu \gamma)$ versus ${\hat F}$; 
the curves  $(a)$ and $(b)$ refer 
to the values: $\Gamma(D^{*0} \to D^0 \gamma)=22 \; KeV$ and 
$\Gamma(D^{*0} \to D^0 \gamma)=11 \; KeV$, respectively.} 
\label{th251_fig3}
\end{figure}

The determination of $f_B$ from
$B^- \to \mu^- {\bar \nu}_\mu \gamma $ relies on heavy quark symmetries. In 
fact, in the  limit $m_b \to \infty$  
$f_B$ can be related to the $B^*$ decay constant  $f_{B^*}$,
defined by: $<0|{\bar b} \gamma_\mu q|B^*(p, 
\epsilon)>=f_{B^*} m_{B^*} \epsilon_\mu$, by the 
relation: $f_B=f_{B^*}={{\hat F} \over \sqrt{m_b}}$, with ${\hat F}$  
independent of $m_b$. Theoretical analyses 
\cite{neubert} indicate: ${\hat F} \simeq 0.35 \; 
GeV^{3/2}$. 
To relate $B^- \to \mu^- {\bar \nu}_\mu \gamma$ 
to  ${\hat F}$, let us 
consider that the decay proceeds through bremsstrahalung and 
structure dependent diagrams. 
\begin{figure}[ht]
\begin{center}
\input FEYNMAN
\begin{picture}(20000,2000)
\THICKLINES
\drawline\fermion[\E\REG](0,0)[2000]
\global\advance\pmidx by -500
\global\advance\pmidy by -900
\put(\pmidx,\pmidy){$\st B^-$}
\put(2000,0){\circle*{400}}
\drawline\photon[\N\REG](\pbackx,\pbacky)[2]
\global\advance\pmidx by 700
\global\advance\pmidy by 200
\put(\pmidx,\pmidy){$ \st \gamma$}
\drawline\fermion[\E\REG](\pfrontx,\pfronty)[2000]
\global\advance\pmidx by -500
\global\advance\pmidy by -900
\put(\pmidx,\pmidy){$\st B^{*-}$}
\put(\pbackx,\pbacky){\rule[-0.6mm]{1.2mm}{1.2mm}}
\drawline\photon[\E\REG](\pbackx,\pbacky)[2]
\drawline\fermion[\NE\REG](\photonbackx,\photonbacky)[2000]
\global\advance\pmidx by -300
\global\advance\pmidy by 500
\put(\pmidx,\pmidy){$\st \mu^-$}
\drawline\fermion[\SE\REG](\photonbackx,\photonbacky)[2000]
\global\advance\pmidx by -200
\global\advance\pmidy by -900
\put(\pmidx,\pmidy){$\st {\bar \nu}$}

\drawline\fermion[\E\REG](10000,0)[2000]
\global\advance\pmidx by -500
\global\advance\pmidy by -900
\put(\pmidx,\pmidy){$\st B^-$}
\put(12000,0){\circle*{400}}
\drawline\photon[\N\REG](\pbackx,\pbacky)[2]
\global\advance\pmidx by 700
\global\advance\pmidy by 200
\put(\pmidx,\pmidy){$ \st \gamma$}
\drawline\fermion[\E\REG](\pfrontx,\pfronty)[2000]
\global\advance\pmidx by -500
\global\advance\pmidy by -900
\put(\pmidx,\pmidy){$\st B^{\prime -}_1$}
\put(\pbackx,\pbacky){\rule[-0.6mm]{1.2mm}{1.2mm}}
\drawline\photon[\E\REG](\pbackx,\pbacky)[2]
\drawline\fermion[\NE\REG](\photonbackx,\photonbacky)[2000]
\global\advance\pmidx by -300
\global\advance\pmidy by 500
\put(\pmidx,\pmidy){$\st \mu^-$}
\drawline\fermion[\SE\REG](\photonbackx,\photonbacky)[2000]
\global\advance\pmidx by -200
\global\advance\pmidy by -900
\put(\pmidx,\pmidy){$\st {\bar \nu}$}
\end{picture}
\end{center}\vskip 0.6cm
\caption{Polar diagrams contributing to $B^- \to \mu^- {\bar \nu}_\mu \gamma$ 
\label{diag}}
\end{figure}
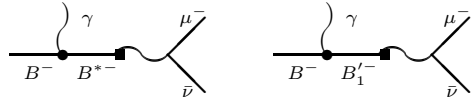
\noindent
The former vanish if one puts $m_\mu=0$, 
 as we do; the latter are polar diagrams (Fig. \ref{diag}), the 
pole being  either a $B^*$ or a $B_1^\prime$; this second state is the only 
$J^P=1^+$ meson which couples
to the weak current in the $m_b \to \infty$ limit. 
We get \cite{noi}:
\begin{eqnarray}
 \Gamma(B^- &\to& \mu^- {\bar \nu}_\mu \gamma)  = 
\Gamma^{(B^*)} + \Gamma^{(B^\prime_1)} 
\nonumber \\
&=&    \int_0^{{m_B \over 2}} 
{2 \; dE_\gamma \; E_\gamma^3 (m_B-2E_\gamma) \over 3 (2\pi)^3} 
\Bigg[ {|C_1|^2 f_{B^*}^2 \over (E_\gamma+\Delta)^2 } +
{|C_2|^2  f_{B^{\prime}_1}^2 \over 
(E_\gamma+  \Delta^{\prime})^2 } \Bigg]
\; ,
\label{width} 
\end{eqnarray}
\noindent
where $C_1$ ($C_2$) depends on the coupling $g_{B^* B\gamma}$ 
($g_{B^\prime_1 B\gamma}$)
 \cite{noi}. Expression (\ref{width}) gives 
access to $f_{B^*}$, and hence to ${\hat F}$, if 
$\Gamma^{(B^\prime_1)} \ll \Gamma^{(B^*)}$. 
Using theoretical inputs to compute 
(\ref{width}), we found: 
$\Gamma^{(B_1^\prime)}\simeq \Gamma^{(B^*)}/10$, 
concluding that the contribution of the second diagram can be included in the 
uncertainty on $f_B$.
Within this approximation, $\Gamma(B^- \to \mu^- {\bar \nu}_\mu 
\gamma)$ is proportional to ${\hat F}^2$. We propose a method  to get
${\hat F}$ based on the experimental knowledge of both 
${\cal B}(B^- \to \mu^- {\bar \nu}_\mu \gamma)$
and $g_{B^* B \gamma}$. 
The prediction
${\cal B}(B^- \to \mu^- {\bar \nu}_\mu \gamma) \simeq {\cal O}(10^{-6})$ 
suggests that this measurement could be accessible at the future $B$ factories 
\footnote{A recent analysis \cite{artuso} gives: 
${\cal B}(B \to \mu \nu_\mu \gamma)<9.5 \cdot 10^{-5}$;
${\cal B}(B \to e \nu_e \gamma)<1.6 \cdot 10^{-4}$.}.
As for $g_{B^* B \gamma}$, it is not
experimentally known, but heavy quark symmetries relate it
to the analogous $D$ coupling $g_{D^* D \gamma}$ \cite{noi2}. 
The measurement: 
${\cal B}(D^{*0} \to D^0 \gamma)=36.4 \pm 2.3 \pm 3.3 \%$ is already available
\cite{cleo1}. 
 Since the bound $\Gamma(D^{*0})<131 \; KeV$ \cite{accmor} 
is not far from current predictions \cite{noi2}, it is possible that 
$\Gamma(D^{*0})$ will be measured in the next future.
Such measurement  would give $g_{D^* D \gamma}$. To show the sensitivity of the 
method, we used two theoretical inputs
for $g_{D^* D \gamma}$. The plot of 
${\cal B}(B^- \to \mu^- {\bar \nu}_\mu \gamma)$ versus ${\hat F}$ is shown in 
Fig. \ref{th251_fig3} \cite{noi}.
The experimental result for 
${\cal B}(B^- \to \mu^- {\bar \nu}_\mu \gamma)$ would give ${\hat F}$. 
In the expected range of values of ${\hat F}$, the results in
Fig. \ref{th251_fig3} confirm that 
${\cal B}(B^- \to \mu^- {\bar \nu}_\mu \gamma)$ should  be ${\cal O}(
10^{-6})$.
\\

\section*{Acknowledgments}
I thank P. Colangelo and G. Nardulli for their collaboration.

\end{document}